\documentclass[
 preprint,
 superscriptaddress,
 amsmath,amssymb,
 aps,prb
]{revtex4-2}
\usepackage{graphicx}
\usepackage{dcolumn}
\usepackage{bm}

\usepackage{multirow}
\usepackage[table,xcdraw]{xcolor}
\usepackage{colortbl}

\begin{document}

\title{Thermal Corrections and Analysis on the Phase Stability of $\boldsymbol{CsPbCl_3}$ and $\boldsymbol{Cs_2AgSbCl_6}$ during In-Situ Thermal Treatment}

\author{Ethan R. Cronk}
\affiliation{Department of Physics and Astronomy, University of Maine, Orono, Maine, 04469, USA}
\affiliation{Frontier Institute for Research in Sensor Technologies, University of Maine, Orono, Maine, 04469, USA}

\author{Wenjun Xiang}
\affiliation{School for Engineering of Matter, Transport and Energy , Arizona State University, Tempe, Arizona, 85281, USA}

\author{Rachel Fister}
\affiliation{Department of Physics and Astronomy, University of Maine, Orono, Maine, 04469, USA}
\affiliation{Frontier Institute for Research in Sensor Technologies, University of Maine, Orono, Maine, 04469, USA}

\author{Biswajit Ball}
\affiliation{Department of Materials Science and Engineering Department, University of Central Florida, Orlando, Florida, 32816, USA}

\author{Feng Yan}
\affiliation{School for Engineering of Matter, Transport and Energy , Arizona State University, Tempe, Arizona, 85281, USA}

\author{Liping Yu}
\affiliation{Department of Materials Science and Engineering Department, University of Central Florida, Orlando, Florida, 32816, USA}

\author{Nicholas S. Bingham}
\affiliation{Department of Physics and Astronomy, University of Maine, Orono, Maine, 04469, USA}
\affiliation{Frontier Institute for Research in Sensor Technologies, University of Maine, Orono, Maine, 04469, USA}

\date{\today}

\begin{abstract}
Cesium lead chloride ($\mathrm{CsPbCl_3}$) is a well known and principal model for inorganic perovskite halide optoelectronic research. The many available techniques including high temperature stability testing have been used to investigate the increasing interest in inorganic perovskites as primary layers in solar cell applications. Due to the nature of high temperature testing, the characterization technique, reproducibility, and the true sample temperature are vital in determining relative stability. By choosing $\mathrm{CsPbCl_3}$ in the investigation of the structural stability of perovskites at high temperatures, it acts as a baseline to create and verify a methodology that accurately probes sample temperature, phase transitions, and decomposition onsets. Therefore, we present a methodological approach to investigate the thermal interactions and stability of $\mathrm{CsPbCl_3}$ as a parent single perovskite halide based on ligand-assisted re-precipitation synthesis techniques. Where we use our approach to inform and probe thermal interactions in other cesium/chlorine compounds like $Cs_2AgSbCl_6$.  By analyzing the stoichiometry and initial phases through investigations of the crystalline structure and particle morphology,  we calculated temperature conversions using a control substrate and refinements to best estimate changing structure parameters. Using in-situ temperature dependent X-ray diffraction, we were able to effectively probe the phase transitions and decomposition temperatures of the investigated halide powders. Creating a process that can confirm known high temperature structural phenomena of model perovskite halides while verifying our true sample temperature. Which allowed for further testing on the thermal kinetics of on the double perovskite structure $Cs_2AgSbCl_6$ and will continue to allow us to test other perovskites and perovskite families of interest in modern high temperature perovskite halide research. 
\end{abstract}

\maketitle


\section{\label{sec:level1} Introduction}
In the study of next generation photoelectronic devices, alternative materials such as metal halide perovskites have been of high interest due to their outstanding optoelectronic properties \cite{Wang2021ProspectsCells, Haque2020HalideThermoelectricity, Manser2016IntriguingPerovskites} and the volume of elementally diverse materials that fit the base perovskite structure \cite{Li2020PressureMorphologies, Simenas2024PhasePerovskites}. 
$CsPbCl_3$ can be seen as one of the central models of this five atom structure and can act as a primary baseline in comparison of perovskite materials for stable  photovoltaic devices  \cite{You2023TheProperties, Aktary2022AApplications}.  The $PbX_6$ octahedra in the corners of the unit cell allow for various phases based on the displaced angles between the divalent Pb cation and the X anion \cite{Wang2021ProspectsCells, Zhang2024RecentNanomaterials}. This displacement along with the neutral (zero) total charge allows for the commonly seen orthorhombic $(\gamma)$, tetragonal $(\beta)$, and cubic $(\alpha)$ phases depending on the formation energies, with the gamma phase being the most favorable due to its lowest ground state energy \cite{Rosales2023LeveragingPerovskites,Jin2021CanPerovskites, Fransson2023RevealingMAPbI3, Bechtel2018First-principlesSolutions}. In addition to the differing phases, these single halide perovskites have a high amenability for various elemental substitutions. \cite{Travis2016OnSystem, Bartel2019NewHalides, Dawa2024ExploringReview}. 
As a result, perovskite halides have been subject to numerous studies investigating the viability of these elementally diverse materials in photovoltaics \cite{Klug2017TailoringProperties, Wang2022IScienceComponents, Bartel2019NewHalides}.  \\
An example of this type of doping, especially when comparing dopants on a specific site are  double perovskite structures. First investigated when exploring alternative ferroelectric materials\cite{Morss1970PreparationCs2NaMCl6}, double perovskite structures have shifted focus to a potential stable structure for non toxic elements and wider band gaps for inorganic optoelectronic integration\cite{Grandhi2024Wide-BandgapOptoelectronics, Zhou2023Wide-bandgapPhotodetectors}. $Cs_2AgSbCl_6$, a double perovskite structure specifically investigated due to its lack of Pb\cite{Zhou2023Wide-bandgapPhotodetectors, Zhou2018ExploringApproach} and chemically unstable Sn\cite{Milot2018TheFilms, Ricciarelli2020InstabilityOxidation,Zhou2018ExploringApproach}, acts as an ideal perovskite in which to probe the effects of elemental substitutions on a specific site (in this case the B-site when compared to $CsPbCl_3$).\\
A major deciding factor in determining whether a material  is suitable for photovoltaic application is stability at high temperatures. Devices like solar cells require negligible loss of photovoltaic performance at $85 ^\circ$C \cite{Mazumdar2021StabilityRemedies} and can reach over $100 ^\circ$C in satellite integration \cite{Meneses-Rodriguez2005PhotovoltaicTemperatures},  so stability within the primary photo-layers at higher temperatures is essential.  Thus, an investigation that probes the dependence at high temperature is necessary for realistic optoelectronic integrations and the choice of elemental substitution.  The research involved in investigating the stability of the structures of perovskite halides at high temperatures can vary depending on the systematic setup or device. Therefore, in-situ diffractometry, structure dependence, or elemental change in perovskite halides require comprehensive synthesis techniques and methodologies to accurately study these properties. In our study on the thermal interaction and stability of perovskite halides, we present a practical and effective methodology in our approach for accurate sample stability, phase transitions, and decomposition temperatures that we compare to similar studies of $CsPbCl3$ and $Cs_2AgSbCl_6$ if available.  To allow for investigations on alternative phase or doped perovskites, studying perovskite materials with known phase phenomena and amenability to elemental dopants allows us to select a synthesis technique that best fits this purpose. Techniques such as vapor deposition and solvent precipitation are well documented as effective approaches for the synthesis of perovskite halides \cite{Costa2017OnMethod, Zhao2016Organic-inorganicApplications}, but it is also seen that the choice of factors in the synthesis processes can affect the initial phases of perovskites, such as temperature, time, solvents used, and starting states \cite{Kayalvizhi2022HydrothermalMechanism,Song2015ImpactPerovskites}.  Ligand-assisted re-precipitation (LARP) is a powerful technique for its ability to synthesize 3D perovskite polycrystalline powders \cite{Sun2023TowardNanocrystals, Xiang2024DoubleCells}, but also for its success seen in successfully doping perovskite halides \cite{Kim2025UnderstandingNanocrystals, Fatemi2024CaCl2Synthesis}. The ability to synthesize various perovskite materials in a comparable way allows for the implementation of techniques that are involved in characterizing the thermal stability of inorganic perovskite halides. Not only to probe the properties and confirm model values in single perovskite halides, but also to conduct future investigations into the effects of further elemental substitutions or doping of these perovskite structures. \\
Therefore, in this research, we present our investigation into our methodology to test, reproduce, confirm, and present known phase transitions and emergent structure decompositions of $CsPbCl_3$ and the not yet reported structure evolution and possible decomposition products and their structures of $Cs_2AgSbCl_6$ as we vary temperature. The results authenticating our methodology for further investigation of the thermal stability of alternative perovskites in optoelectronic application. 
\section{Methods}
Two samples were prepared, $CsPbCl_3$, and $Cs_2AgSbCl_6$ \cite{Yandri2022PhotoluminescenceAnti-solvents}. The initial stoichiometric compounds were combined: Cesium Chloride ($CsCl$) [99.9\%, Thermal Scientific Chemicals] and Lead Chloride ($PbCl_2$) [99.9\%, Thermo Scientific Chemicals] for $CsPbCl_3$. 
Cesium Chloride ($CsCl$) [99.9\%,BeanTown Chemical], Silver Chloride ($AgCl$) [99.995\%, BeanTown Chemical], and Antimony trichloride  ($SbCl_3$) [99\%, BeanTown Chemical] for $Cs_2AgSbCl_6$.
All were used as received without further purification. The three compounds were mixed into a polar solvent Dimethyl sulfoxide (DMSO, purchased from Sigma-Aldrich) to form a precursor of 0.1 Mol. Each desired perovskite halide was then mixed with an anti-solvent Tolunene (99.5\%, Sigma Aldrich). Vigorously stirring in air at room temperature for 24 hours, the homogeneous solution evaporates the anti-solvent, forming our polycrystalline powder. 
For characterization techniques like Scanning Electron Microscopy (SEM) and X-ray Photoelectric Spectroscopy (XPS), a polycrystalline powder form is suitable for characterization, but for our X-ray Diffractometer (XRD, PANalytical X'pert Pro), this is not the case. For non-grazing incident geometries, the XRD has a sample stage that is perpendicular to the ground and does not allow for free standing powder to be scanned while being thermally tested. For this, we utilized a simple drip coating technique to adhere our powders to Si (100) substrates to allow adequate mounting, signal, penetration, and orientation.\\
First, the Si (100) (purchased from University Wafer) substrate surface was cleaned by washing the wafer in Acetone (99.5\%, Fisher Chemicals), Isopropanol (IPA, 99.5\%, Fisher Chemicals), Methanol (99.9\%, Fisher Chemicals), and DI water in a sonic bath (in sequential order) for 5 minutes each. Once cleaned, 5 mg of the compound was mixed with 5 mL of IPA. Using a pipette, the powder suspended in the IPA is drop casted onto the cleaned Si (100) wafer. Once the substrate is sufficiently covered, we wait until the IPA has fully evaporated while the sample is at room temperature. We then have a powder electrostatically bonded to our substrate for temperature varying XRD analysis. This step is repeated for each Perovskite Halide. \\
A Zeiss NVision 40 FIB SEM with an attached Genesis EDS detector is used for imaging and elemental composition confirmation.\\
XPS spectra were obtained using a SPECS  Phoibos hemispherical analyzer under vacuum ranging from \begin{math}6\cdot10^{-9}\end{math} to \begin{math}5\cdot10^{-8}\end{math} Torr. All compounds were scanned before thermal treatment at a step size of 0.1 eV, a dwell of 0.2 seconds, and a pass energy of 20 eV over 15 scans for resolute intensity values. CasaXPS was used for the post scan analysis. \\
Pre-annealed XRD spectra and in-situ crystal phase information was taken with the Malvern Panalytical X'pert 3 Pro MRD diffractometer with 1.54 Å Cu K$\alpha$ radiation through the incident mirror optic and received by the Pixcel 3 detector. We characterized the room-temperature structure of the powders by mounting each on tape affixed to amorphous glass slides for pre-annealed phase confirmation.\\
The Anton-Parr dome heating stage (DHS 1100) is an attachment to the Panalytical diffractometer sample stage with a graphite dome to ensure stability in environmental temperature around the sample and to allow for inert atmospheres like $N_2$. The dome is flushed of air with a flow of 225 Torr of $N_2$ until at equilibrium pressure. Because of the geometry of the graphite dome and the attenuation of the x-rays a percentage of the diffracted beam off of the dome can still be sensed by the diffractometer. To counteract this, we attach a plastic fabricated shield to allow sample intensity but block most of the dome intensity \cite{VanDerPers2021ADiffractometers}. \\
\section{Domed Heating Stage Attachment Calibration}
A large part of the process of thermally testing these perovskite halides comes from calibrating the temperatures of our synthesized powders dropcasted on \textit{Si} (100) substrates. The Anton Paar DHS 1100 is composed of the resistance based kanthal heater and 1 mm thick Aluminum Nitride (AlN) heating plate, where our Si substrate sample is then affixed to the AlN heating plate with inconel clips. This layering of the DHS heating components under our sample allows for various points of temperature loss through the thermal conductance of each layer and the heat sinks of the clips. In this way, with the k-type thermocouple reading the internal temperatures of the DHS lying just below the kanthal heater, we lose the specificity of the true temperature of the sample powder when the temperature is raised. To counteract this, we performed  various calibrations to find substrate thermal expansion constants to correctly estimate the temperature difference between the thermocouple and the actual sample temperature.
Thus, in a $N_2$ atmosphere, we calibrated our \textit{Si} (100) substrate to 800\begin{math}^\circ\end{math}C to have a full characterization of the temperature difference and, in turn, the sample temperature. Detecting the \textit{Si} (400) peak at every 100\begin{math}^\circ\end{math}C from room temperature to 100\begin{math}^\circ\end{math}C then up to 800\begin{math}^\circ\end{math}C we find the $2\theta$ position at each step. To calculate the d-spacing along the crystallographic a-axis of the \textit{Si} (100) substrate, we take our $2\theta$ positions at each temperature and use equation (1).
\begin{equation}
d_{hkl} = \frac{n\lambda}{2sin\theta}
\end{equation}
Then apply each d-spacing value to get the percentage of the a-axis lattice parameter for a cubic crystal using, 
\begin{equation}
a = d_{hkl}\sqrt{h^2+k^2+l^2} 
\end{equation}

With these values, we can calculate the actual sample temperature using the thermal expansion equation with known thermal expansion coefficients for Si \cite{TouloukianY1977ThermophysicalSolids.} to get Figure 1. 
\begin{equation}
\frac{\Delta L}{L[T]}[\%] = a_o +a_1\cdot T + a_2\cdot T^2 + a_3\cdot T^3  
\end{equation}
with,
\begin{equation}\Delta L = L(T) - L(T_I)\end{equation}where\\
L(T)\hspace{1.4cm} Length at temperature T in Angstroms\\
T \hspace{1.8cm} Temperature in K\\
\begin{math}T_I\end{math} \hspace{1.75cm} Initial temperature in K\\
\begin{math}a_0, a_1, a_2, a_3\end{math} \hspace{.01cm} Thermal expansion coefficients.\\

\begin{figure}[h]
    \centering
    \includegraphics[width=.75\linewidth]{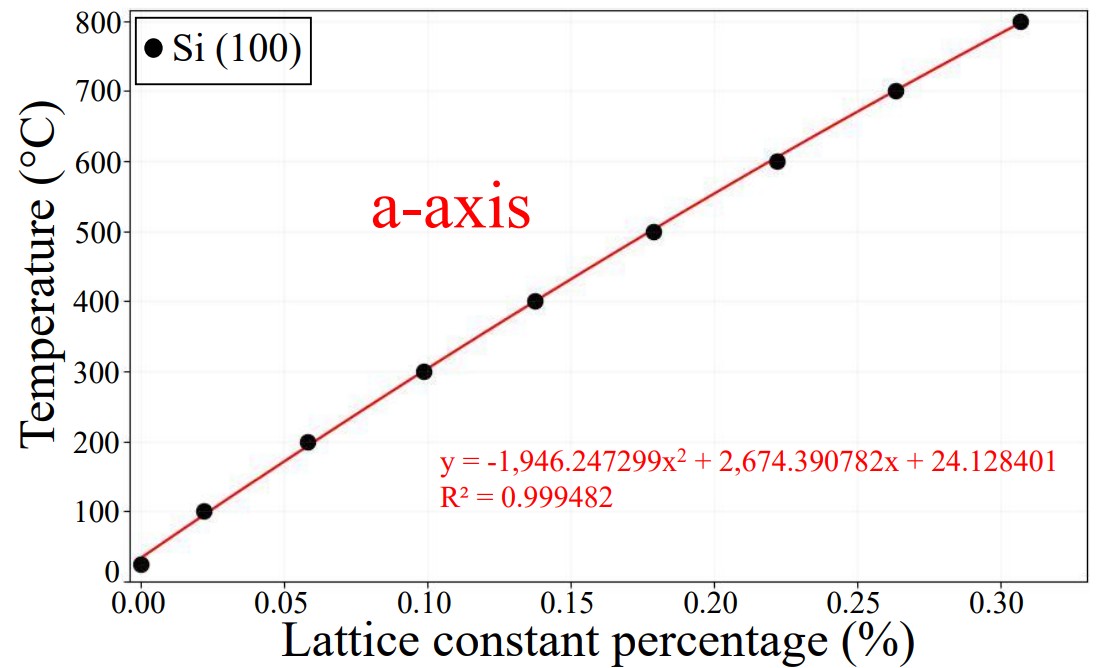}
    \caption{Si (100) substrate temperature conversion equation is calculated and fitted. Constant percentages were calculated on the crystallographic a-axis from room temperature to 800\begin{math}^\circ\end{math}C. Giving a statistically significant $R^2$ near one.}
\end{figure}
Using the lattice constant percentage calculated at each temperature from the Si reference, fitting to a third- or second order polynomial \cite{AntonPaarThermalXRD}, we inserted our values into the fit in Figure 1 to obtain the temperature change between the thermal couple and our sample. Then by flipping the axes of our calculated LCP vs. temperature we can convert any temperature step to get a Si lattice constant percentage. Using the literature lattice constant percentage fit to get the conversion for any temperature value. \\
After taking into account the temperature conversions to probe realistic sample temperatures, we investigate the characteristics of each chosen material prior to in-situ thermal testing. Here we confirm consistent and homogeneous material stoichiometry and beginning phase to allow us to correctly compare our material with other known values to confirm the process when prepping and using in-situ XRD for structure analysis. 

\section{Results: Single Perovskite Structure}
\subsection{Stoichiometric Characterization}
\begin{figure}[h]
    \centering
    \includegraphics[width=1\linewidth]{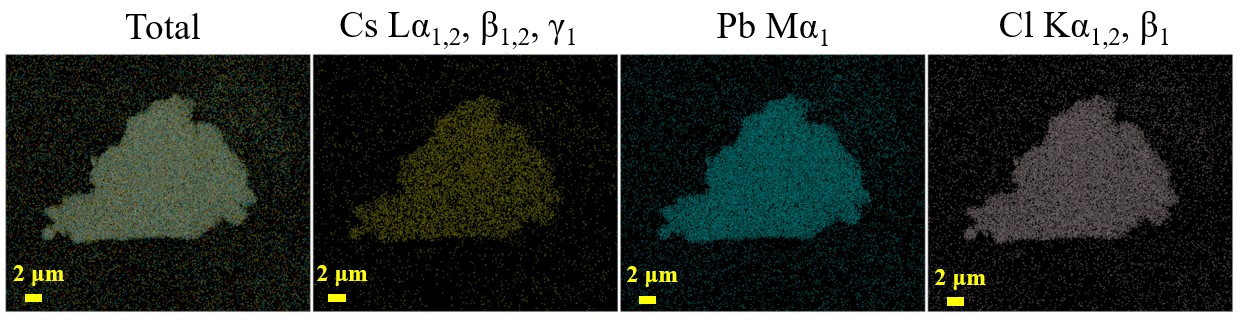}
    \caption{\textit{Atomic ratio and homogeneity is probed using EDS on a single grain of $CsPbCl_3$. Two scans were taken at an acceleration voltage of 15 kV using a working distance of 5 mm at 3250X magnification. }}
\end{figure}
Fig.2 displays EDS elemental intensity images before annealing for the single inorganic perovskite halide $CsPbCl_3$. It is observed in Fig.2 that prior to thermal treatment we obtain clear signal from the A, B and X-site elements and a smooth and homogeneous distribution from their corresponding L, K, and M x-ray absorption edges (SEM image of the EDS scan listed in Figure S1). To further confirm synthesis, we compare our experimental atomic percentages with the known atomic ratio for the \begin{math}ABX_3\end{math} structure (20:20:60). It is necessary to confirm these ratios for the relative homogeneity we expect post synthesis. To obtain the atomic ratios, Genisis EDS uses the peak to background ratios of the elemental signal along with iterative ZAF correction values to calculate the weight percentages. Then by dividing the weight percentage by each element's atomic weight and dividing that value by the sum of each element value,\begin{equation}\frac{Wt(\%)}{At. Wt.}\\\end{equation}
we obtain our atomic percentages. In Table I the atomic percentages lie close to the ideal ratio within the standard error of the standardless ZAF iteration method \cite{Eggert2020Effect_of_the_silicon_drift_detector_on_edax_standardless_quant_methods}. \\

\begin{figure}[h]
    \centering
    \includegraphics[width=1\linewidth, keepaspectratio]{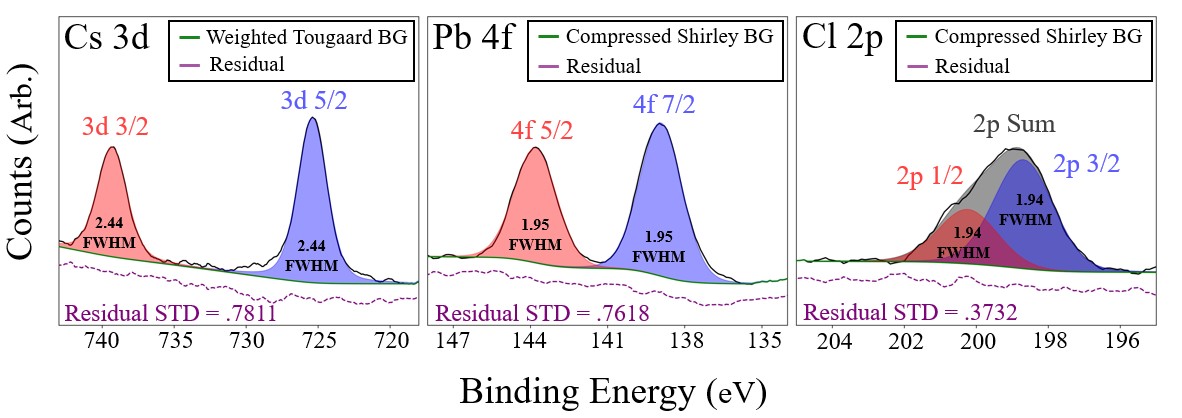}
    \caption{\textit{XPS spectra is shown for the characteristic A,B, and X element peaks for $CsPbCl_3$. The spectrum is charge corrected using the C 1s positioning before analysis and the standard deviation is calculated based on the residuals of the CasaXPS models.}}
\end{figure}
In addition to exploring the stoichiometry of our samples, we approach the materials in a similar fashion by investigating the surface of our powders with XPS to determine clear signal and atomic percentages.  In Fig. 3 we observe similar results as in Fig. 2, obtaining an adequate signal from each elemental component for $CsPbCl_3$. To best calculate the atomic percentages for comparison, background estimations and corrections were used on the basis of the element and signal.  Where the weighted Tougaard best estimated the pre-corrected inelastic scattering seen in the background of the Cs 3d orbital pairing and the compressed Shirley estimation best fit the background rising with the expected core signals for Pb 4f and Cl 2p.  Alongside the background, the peak shape estimation was restricted to (1) each spin-orbit couple was fit to a shared FWHM value, (2) every orbital was matched with its corresponding relative sensitivity factors for the atomic percentage calculation based on signal intensities, and (3) peak shapes were modeled using a percentage distribution of a Lorentzian and Gaussian fit to best match each orbital shape and any asymmetry. The residual standard deviations are shown to best visualize effective models.   
When comparing the calculated atomic ratios from the EDS and XPS in Table I, we see an agreement, within error, between the two techniques. Confirming that our powders are within the standard in homogeneity and stoichiometry of the \begin{math}ABX_3\end{math} structure for $CsPbCl_3$. Where some difference can be attributed to possible halide segregation or carbonates on the surface due to ambient environmental degradation pathways \cite{Chen2025RefiningCells, DeKeersmaecker2025ActivatedSpectroscopy}.
\begin{table}[h!]
    \centering
    \caption{ Atomic ratio of $\mathrm{CsPbCl_3}$. The total ratios are compared between EDS and XPS characterization where the significance of each individual orbital in XPS is shown. The XPS error is calculated based on the total peak area models done in CasaXPS.}
        \centering
     $\mathrm{CsPbCl_3}$ \\[6pt]
        \begin{tabular}{ccccc}
            \hline
            Element & Device & Orbital & \multicolumn{2}{c}{Atomic Percentage(\%)}\\
            \hline
            & EDS & -- & -- & 20.10 $\pm$ 10\\
               Cs& \multirow{2}{*}{XPS}& 3d\,5/2 & 8.26 & \multirow{2}{*}{$16.06 \pm 3.97$}\\
               &     & 3d\,3/2 & 7.80 & \\[6pt]
            & EDS & -- & -- & 21.83 $\pm$ 10\\
               Pb& \multirow{2}{*}{XPS}& 4f\,7/2 & 11.29 & \multirow{2}{*}{$22.02 \pm 2.85$}\\
               &     & 4f\,5/2 & 10.73 & \\[6pt]
            & EDS & -- & -- & 58.07 $\pm$ 10\\
               Cl& \multirow{2}{*}{XPS}& 2p\,3/2 & 31.38 & \multirow{2}{*}{$61.92 \pm 7.78$}\\
               &     & 2p\,1/2 & 30.54 & \\
            \hline
        \end{tabular}

\end{table}

\subsection{Structure Characterization}
For $CsPbCl_3$, an as grown XRD is taken at room temperature to determine the initial phase and confirm how our synthesis affected the inherent ground state energy if at all. Mounting the powders of each material, taken from the same synthesized batch, on amorphous tape on top of glass slides,  10-35\begin{math}^\circ\end{math} \begin{math}2\theta\end{math} XRD scans were taken with the same optics as our in-situ tests. Rietveld refinements were performed using Panalytical's Highscore Plus to run the simulations based on our diffractometer specifications and best structure matches in the ICDD PDF Powder Diffraction database [Fig 4]. Based on the best PDF fits, we see a shared starting orthorhombic phase $(\gamma)$. This confirms that we are not straining or forcing any higher formation energies at room temperature due to our choice in synthesis. Where it is commonly seen in LARP methods using DMSO and Toluene that 3D bulk nano-crystalline powders left over after stirring can form the $(\gamma)$ phase \cite{Bertolotti2017CoherentNanocrystals, Tsvetkov2020FormationMaterials, Pal2018All-Solid-StateHalides}. \\

\begin{figure}[h]
    \centering
    \includegraphics[width=.75\linewidth]{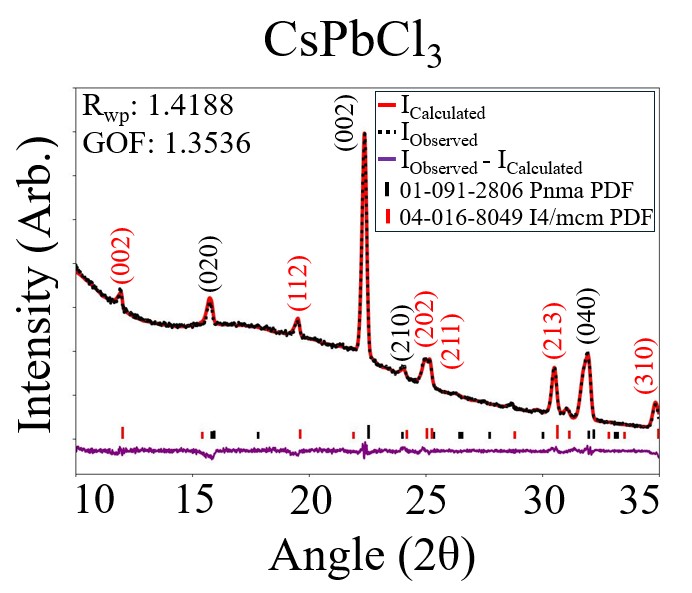}
    \caption{\textit{A pre-annealed XRD spectrum was taken on$\mathrm{CsPbCl_3}$ and then Reitveld Refined to confirm the starting phase. Here, each peak is matched to its corresponding Miller plane defined by the best ICDD PDF match. Each refinement is shown with the residual between the calculated and observed spectra matched with the calculated goodness of fit and weighted R profile}}
\end{figure}

As a note, due to this solvent based synthesis method, it is seen specifically in $CsPbCl_3$ that there can be a dual phase at room temperature. Indicated in the literature for the investigation of 2D synthesis methods for $CsPb_{2}Cl_{5}$ \cite{Becerril2023PhotoluminescenceCVD, Wei2024TheUltraviolet,Pal2018All-Solid-StateHalides} the 3D bulk powder $CsPbCl_3$ commonly forms in an orthorhombic phase, but also forms a secondary 2D $CsPb_{2}Cl_{5}$ phase. This is seen in our refinement (Fig.5) where we get a significant $R_{weighted}$ and Goodness of Fit when refined to orthorhombic $CsPbCl_3$ and tetragonal $CsPb_{2}Cl_{5}$. 
\begin{figure}[h]
    \centering
    \includegraphics[width=1\linewidth, , height=.6\textheight, keepaspectratio]{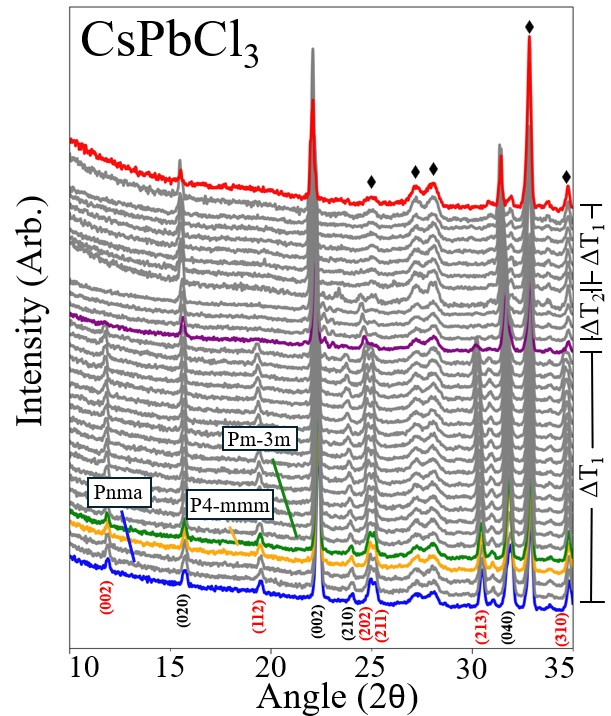}
    \caption{\textit{In-situ XRD scans taken from room temperature to $50 ^\circ C$ under their melting points $\mathrm{CsPbCl_3}$. $\Delta T_1$ signifies a temperature step of $10^\circ$ and $\Delta T_2$ a step of $50^\circ$. The miller planes are listed for each peak corresponding to the material PDF match while the black diamonds represent background peaks due to the ALN heating plate or graphite dome.}}
\end{figure}
\subsection{Thermal Testing}
Using our understanding of the polycrystalline halide powder with the pre-anneal stoichiometry and refinements on the structures, we also investigated the thermal dependence on our initial halide structures. 
To protect our diffractometer and Dome heating stage attachment. We heated $CsPbCl_3$ (a melting point of 615 \begin{math}^\circ\end{math}C \cite{Fayon1996StudyMeasurements})  until we reach the limit of 50 \begin{math}^\circ C\end{math} under their melting points. Heating in 10 degree steps, as read by the thermocouple, at a rate of 2\begin{math}^\circ C\end{math}/min while dwelling at each step for 5 minutes before taking the scan, we can accurately observe the lower temperature phase transitions and gauge a specific onset of decomposition temperature while making sure our sample temperature is stable before running a \begin{math}\theta-2\theta\end{math} scans from 10-35\begin{math}^\circ\end{math}. \\
In our work previous work, \cite{BiswajitBall2026KineticHalides}  we plotted the change in structure as a function of temperature and intensity visualized as a contour plot.  Through this we gauged broad changes in our spectra, whether that be peaks appearing/disappearing, changes in FWHM, or peak splitting. Using this strategy pinpointing specific transitions or onsets of decomposition was difficult to parse, therefore we gained more detail and information from these in-situ XRD scans by refining. We can Rietveld refine to database spectra to get statistical matches to scanned perovskite structures using Panalytical Highscore. We confirmed  starting crystallography with our room temperature spectrum, then applied refinements to every temperature step. Refining each XRD temperature step spectra and calculating various crystal parameters, i.e. D-spacing, volume, refinement statistics, lattice parameters, peak locations, and miller plane matches, we can more specifically find structure transitions and onsets of decomposition. Further detail and alternative examples of this technique can be seen in our work, J. Am. Chem. Soc. 2026, 148, 21, 21694–21703  \cite{BiswajitBall2026KineticHalides}
Using these two plotting strategies we found the close to room temperature orthorhombic to tetragonal phase change at 42.48 $\pm$ 18.302 \begin{math}^\circ\end{math}C  and tetragonal to cubic phase change at 50.97 $\pm$ 18.231 \begin{math}^\circ\end{math}C [denoted in yellow and green in Fig. 5, respectively]. When plotting the the d-spacing vs. temperature we can see a significant slope change in the d-spacing vs. temperature as well[Fig. S4]. After the tetragonal to cubic $CsPbCl_3$  phase transition, at 192.58 $\pm$ 16.990 \begin{math}^\circ\end{math}C we then lose the signal from the $CsPb_{2}Cl_{5}$ dual phase (denoted purple in Fig. 5). 
Lastly, we look to the onsets of decomposition for $CsPbCl_3$ at 465.29 $\pm$ 18.52 \begin{math}^\circ\end{math}C. (shown in red in Fig. 5). Due to the nature of decomposition, these points are chosen to be onsets due to the variability and fluctuations of the $2\theta$ positions and intensity as we approach the full structural decomposition of the ($\alpha$) cubic phase (as seen in Figure 5 and S4). The applied error can be attributed from the sum in quadrature of the measurement error, thermocouple error, and the second-/third order polynomial error used for temperature conversion. Where the measurement drove the most error. In Table. II, when comparing our experimental values to values found in the literature, we see within error that our $CsPbCl_3$ phase transitions are comparable to other studies\nocite{Simenas2024PhasePerovskites, Agrawal2022ThermalPerovskite, Liao2019InNanocrystals}. To note, any structural in-situ decomposition temperatures were not seen in literature. Therefore, we compare this structure decomposition (first shown in \cite{BiswajitBall2026KineticHalides}) with stable up to temperatures from works investigating melting temperatures of $CsPbCl_3$. Our decomposition temperatures come close, but not within the values seen for structures at high temperature in the common literature. However, when comparing our values to these studies, there are some differences. Our synthesis techniques are not exactly analogous to the powders tested in these studies and because of the nature of the synthesis our starting phases also differ. Whether the difference in the onset of structural decomposition is attributed to the lower ground formation energies of these $\gamma$ phase powders, the shared dual phase in 3D $CsPbCl_3$ and 2D $CsPb_{2}Cl_{5}$, or the parameters that are applied for the temperature conversions and in-situ XRD, we believe that any combinations of these possible differences could be an underlying cause. 
\begin{table}[h]
    \centering
    \caption{Phase transitions and structure decomposition values of $\mathrm{CsPbCl_3}$ from this work compared to literature. For decomposition the shown literature examples only list temperatures stable up to since no explicit decomposition was probed\textsuperscript{a}.}
    \begin{tabular}{p{0.23\linewidth}p{0.12\linewidth}p{0.22\linewidth}}
        \hline
        & & $\mathrm{CsPbCl_3}$ (\textdegree C)\\
        \hline
        \multirow{2}{=}{\shortstack[l]{Orthorhombic to \\ Tetragonal Transition}} & This work  & $42.48 \pm 18.302$\\
        & Literature & 41.85\,\textsuperscript{5} \\[6pt]
        \multirow{2}{=}{\shortstack[l]{Tetragonal to \\ Cubic Transition}} & This work  & $50.97 \pm 18.231$\\
        & Literature & 46.85\,\textsuperscript{5} \\[6pt]
        \multirow{2}{=}{\shortstack[l]{Onset of Structure \\ Decomposition}} & This work  & $465.29 \pm 14.254$\\
        & Literature & Stable to 500\,\textsuperscript{49}\\
        \hline
    \end{tabular}
    \begin{minipage}{\linewidth}
        \vspace{4pt}
        \textsuperscript{a} Literature values taken from refs.~\textsuperscript{5,49}.
    \end{minipage}
\end{table}
\subsection{Results: Double Perovskite Structure}
As shown in the single perovskite result section, a pre-anneal EDS scan is done on a single grain on $Cs_2AgSbCl_6$ (Figure 6). Comparing the atomic ratio of all constituent elements we observe: Cs = 19.64\%, Ag = 10.36\%, Sb = 9.91\%, and Cl = 60.10\%. This follows very closely to the expected 20:10:10:60 split we should see in the atomic percentages of double perovskite structures and well within error with the standardless EDS error of $\pm$10\%.
\begin{figure}[h]
    \centering
    \includegraphics[width=1\linewidth]{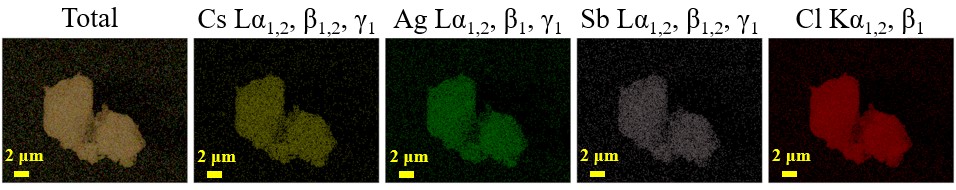}
    \caption{\textit{Atomic ratio and homogeneity is probed using EDS on a single grain of $Cs_2AgSbCl_6$. Two scans were taken at an acceleration voltage of 15 kV using a working distance of 5 mm at 3250X magnification. }}
\end{figure}
After confirming $Cs_2AgSbCl_6$ follows what we would expect for a chemical confirmation of this material, we apply our in-situ testing on $Cs_2AgSbCl_6$ to confirm structure decomposition not yet studied. Due to the hydrothermal synthesis, the equilibrium state at the low formation energy demonstrates a close refined match to a Fm-3m cubic double perovskite structure, which is to be expected\cite{Zhou2018ExploringApproach, Deng2017SynthesisCells} (Fig. 7). Now when investigating the decomposition kinematics of a material like $Cs_2AgSbCl_6$, visualizing the structure in terms of how the miller planes evolve with temperature gives us cues of temperatures of importance. When looking at Figure 8, we see our characteristic cubic peaks stay steady and shift 2$\theta$ due to the material's normal thermal expansion. When comparing to a similar A and X site system of $CsPbCl_3$, where we are probing the effects of substituting differing elements into the B-sites, the apparent loss of structure around 250 $^\circ C$ is quite low. When looking into other studies investigating $Cs_2AgSbCl_6$ as a function of temperature, the closest comparison comes in the form of a Thermogravimetric Analysis (TGA) scan from \cite{Zhou2018ExploringApproach}. Here, the authors claim decomposition temperatures at the two step mass loss of 356 $^\circ C$ and 756$^\circ C$. However, from our XRD measurements (Figure 8) there appears to be a considerable  structure transition much below the primary mass loss of 356 $^\circ C$.  When considering the first derivative of the TGA, around 250 $^\circ C$ we see a slight shift in the spectra as a peak. This could be a sign of structure decomposition comparable to our in-situ scans before material loss. \\
\begin{figure}[ht!]
    \centering
    \includegraphics[width=1\linewidth, , height=.75\textheight, keepaspectratio]{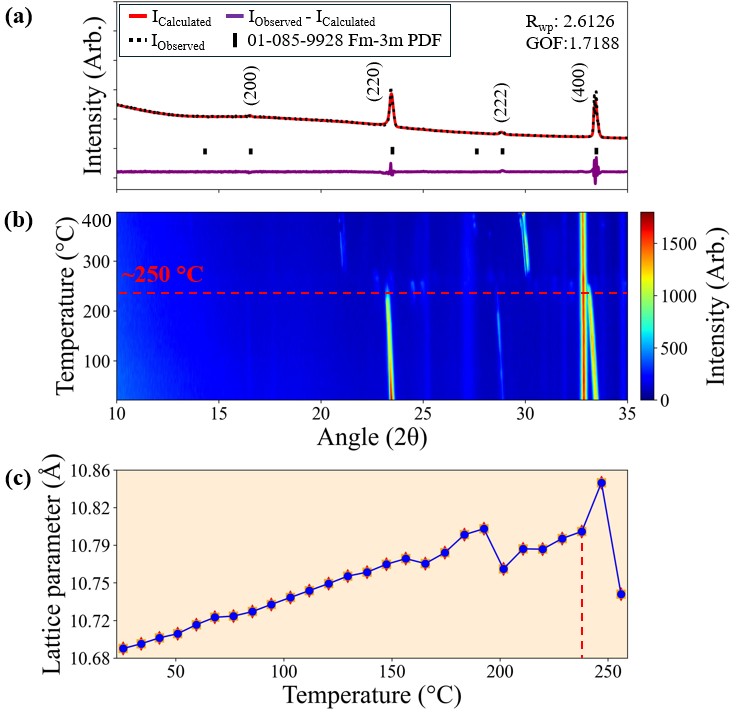}
    \caption{\textit{(a) Pre-annealed XRD spectra were taken on $\mathrm{Cs_2AgSbCl_6}$ and then Reitveld Refined to confirm the starting phase. Here, each peak is matched to its corresponding Miller plane defined by the best ICDD PDF match. Each refinement is shown with the residual between the calculated and observed spectra matched with the calculated goodness of fit and weighted R profile. (b) In-situ XRD scans are shown, taken from room temperature to $400 ^\circ C$ (after conversion) on $\mathrm{Cs_2AgSbCl_6}$. Graphed to track miler planes with first structure transition denoted with the dotted red line. (c) Lattice parameters as a function of temperature are shown for $\mathrm{Cs_2AgSbCl_6}$. The red dashed line denotes the onset of structural decomposition of the primitive unit cell.}}
\end{figure}
Using the same approach as we did for the single perovskite $CsPbCl_3$, we confirm the specific decomposition of $Cs_2AgSbCl_6$ (Fig.7). Starting as Fm-3m cubic (Fig. 7 (a)), we track our planes of interest until we see extreme fluctuations in our major peaks and the lattice parameters (Fig. 8 (b and c)). In Figure 7 (b), by following the (222) and (220) planes we can see at 237.90 $^\circ C$ (denoted in red) we begin our onset of decomposition before full loss of intensity. 
\begin{figure}[ht!]
    \centering
    \includegraphics[width=1\linewidth, , height=.5\textheight, keepaspectratio]{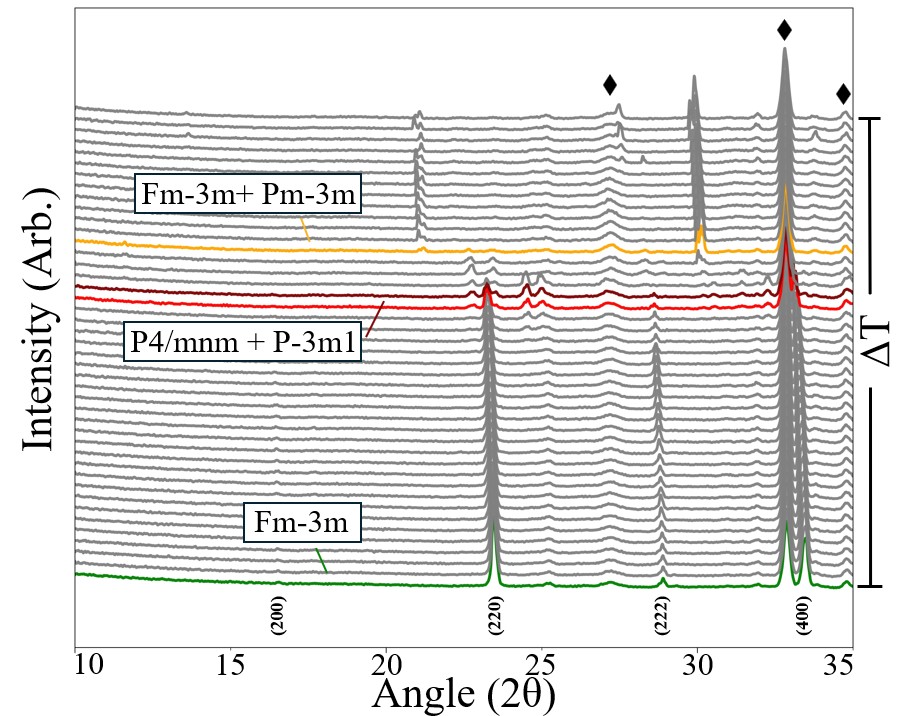}
    \caption{\textit{In-situ XRD scans taken from room temperature to $400 ^\circ C$ (after conversion) on $\mathrm{Cs_2AgSbCl_6}$ . $\Delta T$ signifies a temperature step of $10^\circ$. The miller planes are listed for each peak corresponding to the material PDF match while the black diamonds represent background peaks due to the AlN heating plate or graphite dome.}}
\end{figure}
To note in Zhou et al (2018), the authors cite that the possible decomposition products we should see at each mass loss closely correspond to a separation of compounds due to thermal evaporation that follows as this 

\begin{equation}
Cs_2AgSbCl_6 \rightarrow SbCl_3 + Cs_2AgCl_3   
\end{equation}
\begin{equation} 
Cs_2AgCl_3 \rightarrow AgCl + 2CsCl.  
\end{equation}
Where they based the decomposition on what chemical decomposition would cause a 28.55$\%$ and 42.64$\%$ mass loss. \\
In Figure 8, the full understanding involved with decomposition is yet to been seen, but in addition to the work from Zhou et al. (2018) we can add structure dynamics involved in the multi step decomposition. Right before we see the beginning of the Fm-3m decomposition, various minor peaks appear between 20-30$^\circ$ $2\theta$. When comparing these peaks to known structures in the PDF database we get matches to P4/mnm $CsAgCl_2$ and P-3m1 $Cs_3Sb_2Cl_9$ structures (shown in maroon). Then during the decomposition of these secondary structures we are left with a close match to Fm-3m AgCl and Pm-3m CsCl (shown in orange). Now these matches are PDF structures matched to scans done on specific materials in the database and not exact chemical matches. \\
\begin{figure}[h]
    \centering
    \includegraphics[width=1\linewidth]{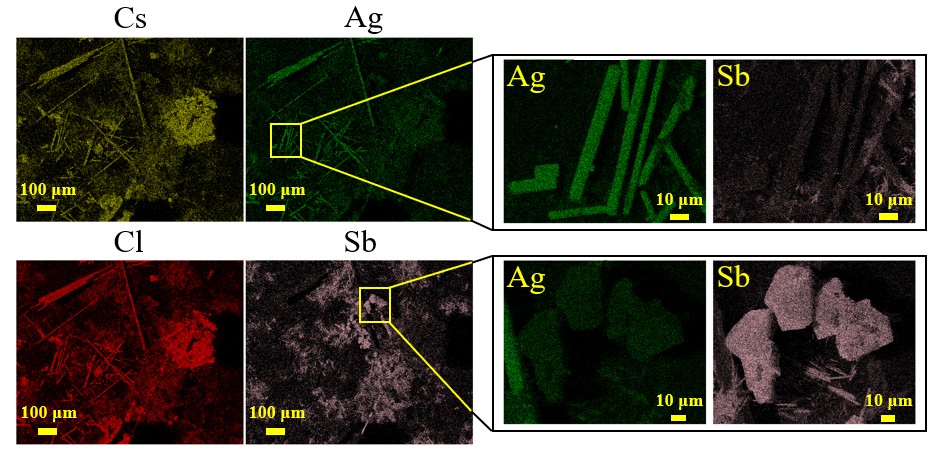}
    \caption{\textit{EDS on $Cs_2AgSbCl_6$ post in-situ heating. A wide spectrum is shown of the post annealed powder. Then two locations of interest are denoted in yellow, where EDS spectra are taken at both spots. Two scans were taken at an acceleration voltage of 15 kV using a working distance of 5 mm. }}
\end{figure}
To further explore possible decomposition products of our material we probed the post in-situ heated powder with EDS. Figure 9 displays a large wide scan EDS of our powder. When exploring the different topographic specimens of interest we found two types of left over material; grains and rods. In Figure 9 we still observe clear Cs and Cl signal throughout, but when honing onto these grains and rods we see a clear differentiation between Ag and Sb. The Ag accumulates in the rods and the Sb in the grains. We zoomed in on a grouping of rods and grains to further verify this B-site element splitting (Figure 9). In the two locations of interest we can still see this elemental separation depending on the differing  forms. There still is signal to be seen from the opposite element, Sb in the Ag dominated rods and Ag in the Sb dominated grains, but this could be due to surface contamination due to elemental migration, thermal evaporation, or decomposition. 
When comparing these post heating EDS results with our in-situ XRD in Figure 8 and the decomposition products stated in Zhou et al. (2018), we see an interesting comparison. We find two defining grain types that correspond to the B-site element segregation we would expect from the decomposition products. Observing the tetragonal and trigonal phase transformation in the in-situ XRD leads to a possible shared $SbCl_3$+ $Cs_2AgCl_3$+$Cs_3Sb_2Cl_9$ multifaceted decomposition before the $CsCl$ and $AgCl$ phase. Where the present Cs, Ag, Sb, and Cl EDS signal corresponds to this intermediate decomposition. These transitions correspond well to the postulated decomposition products seen in Zhou et al. (2018), but the fact that we see these transformations in structure well before the TGA temperature drop offs signifies a possible structure decomposition to the primary cubic $Cs_2AgSbCl_6$ phase and the secondary trigonal and tetragonal phase at a lower temperature before the material fully decomposes near 356 $^\circ C$ and 756$^\circ C$ in the TGA.  
\section{Discussion}
This research, through recreation of the results of the known literature of a single perovskite halide and an investigation on a lesser reported double perovskite structure decomposition, is pertinent due to the variety that can be seen in high temperature testing. Devices varying in specificity, environment, preparation, and synthesis can lead to a vast collection of variables that can affect a single property such as structure decomposition. Due to this, sharing our methodology and results not only adds to this collection of information, it allows for a baseline to apply to further testing. The goal of this testing on well studied perovskites such as $CsPbCl_3$ lies in confirmation. The creation of a process that can be improved and confirms known structural phenomena and properties enables further testing on various materials, such as more heavily substituted and doped halides, like the $Cs_2AgSbCl_6$ shown, or other families such as high entropy perovskites and perovskite oxides. Thus, allowing for in-depth and extensive searches for structural decomposition and how these properties change based on composition. \raggedbottom
\section{Conclusion}
A ligand-assisted re-precipitation technique was utilized to synthesize the inorganic $ CsPbCl_3$ and $Cs_2AgSbCl_6$ perovskite halides. By applying conversions based on substrate temperatures using a lattice constant percentage fitting strategy, we were able to effectively probe the phase transitions and onset of structural decomposition temperatures of the investigated halides. Here we define the relative decomposition as a majority loss of intensity indicating a loss of phase or general structure while under \begin{math}N_2\end{math} atmosphere when annealing. \\
Analyzing the crystal structure of the pre-annealed samples with Rietveld Refinemnet, we confirmed starting orthorhombic phases ($\gamma$) for the 3D  $CsPbCl_3$ halide. It shared a dual phase with the tetragonal 2D $CsPb_{2}Cl_{5}$ given by the noticeable $R_{WP}$ and GOF values. No phase separation and elemental atomic percentages were consistent with the observed \begin{math}ABX_3\end{math} data confirmed by EDS and XPS. However, the pre-anneal analysis of values indicates possible carbonates or halide segregation at the surface hinted by the XPS residuals and EDS percentages. \\
Using the initial phase characterization, we observed phase transitions consistent (within error) with $CsPbCl_3$ found in the literature. Although the onsets of decomposition lie under the known stability temperatures in the literature, a structure decomposition well above 150\begin{math}^\circ\end{math}C is ideal for the upper limit of temperature in uses for most devices in photovoltaics. Comparing the experimental set up from testing in literature, the evidence of a 2D phase and in-situ tests signify a possible earlier onset of structure decomposition. 
Once we confirmed previously noted phase transition temperatures and decompositions we then probed structural stability of a lead free double perovskite structure,  $Cs_2AgSbCl_6$. After confirming beginning phase we were able to present an in-situ look at the Fm-3m structure decomposition and structure transformation of $Cs_2AgSbCl_6$ during and after initial onset of decomposition compared to the literature that was available. Showing a potential structure decomposition prior to material loss when comparing post annealed EDS and the in-situ phase transitions. Overall, this study demonstrates a successful synthesis of 3D perovskite halides, consistent pre analysis of in-situ temperature dependent diffractometry using spectroscopic techniques and refinements of phase, and successful preparation/collection of in-situ x-ray diffraction to probe stability of perovskite halides as we vary temperature. 
\begin{acknowledgments}
This work was supported by the NSF Grant \#2421149 and  in part through the technical support from Dr. Michael Koehler at the Institute for Advanced Materials and Manufacturing (IAMM), University of Tennessee Knoxville. Further support of this research was provided by the facilities at the Frontier Institute for Research in Sensor Technologies (FIRST) and Physics department at the University of Maine. 
\end{acknowledgments}
\bibliography{references}

\clearpage
\onecolumngrid   

\section*{Supplementary Information}

\setcounter{figure}{0}
\setcounter{table}{0}
\renewcommand{\thefigure}{S\arabic{figure}}
\renewcommand{\thetable}{S\arabic{table}}
\renewcommand{\figurename}{Supplementary Figure}

\begin{figure}[ht!]
    \centering
    \includegraphics[width=1\linewidth]{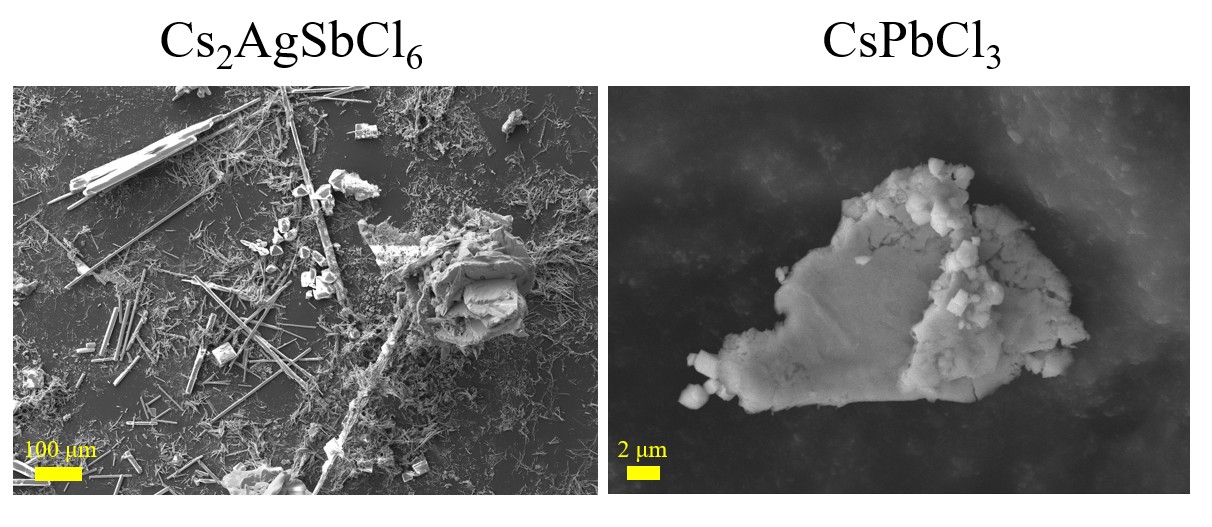}
    
    \caption{SEM images of the single grains of $CsPbBr_3$ and $CsPbCl_3$. These images were taken before EDS spectra was taken.}
\end{figure}

\pagebreak
\begin{figure}[!ht]
    \centering
    \includegraphics[width=1\linewidth]{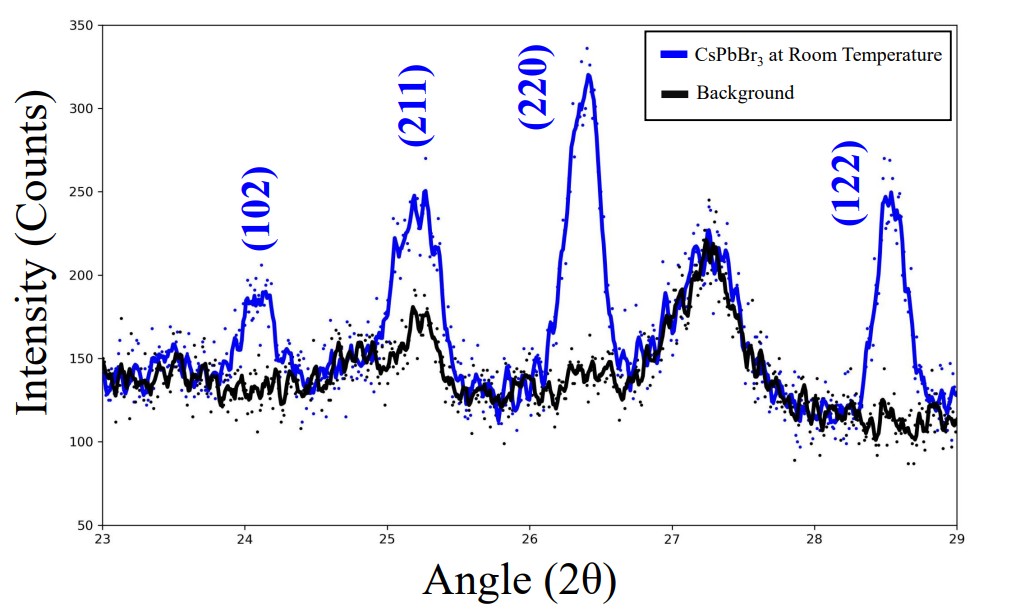}
    
    \caption{A comparison between the room temperature $CsPbBr_3$ XRD spectra and the AlN heating plate and graphite dome signal from the DHS 1100. Where we note that the (221) peak overlaps a background peak. }
\end{figure}
\pagebreak 
\begin{figure}[!ht]
    \centering
    \includegraphics[width=1\linewidth]{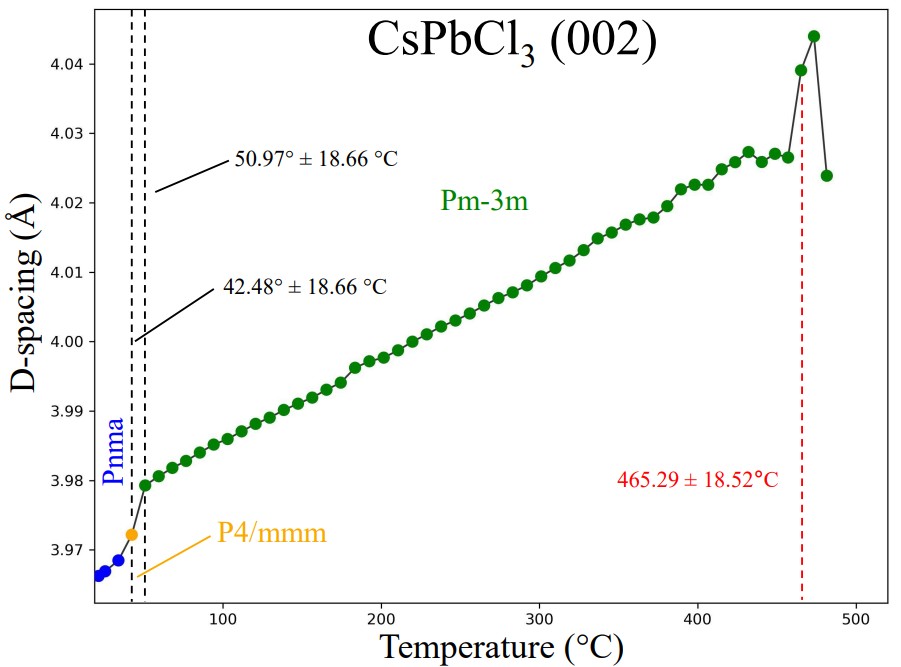}
    
    \caption{D-spacing of the (002) Miller plane of $CsPbCl_3$ taken as a function of temperature. The black dotted lines represent the phase transitions and the red dotted line represents the onset of decomposition.}
\end{figure}
\pagebreak
\begin{figure}[ht!]
    \centering
    \includegraphics[width=1\linewidth]{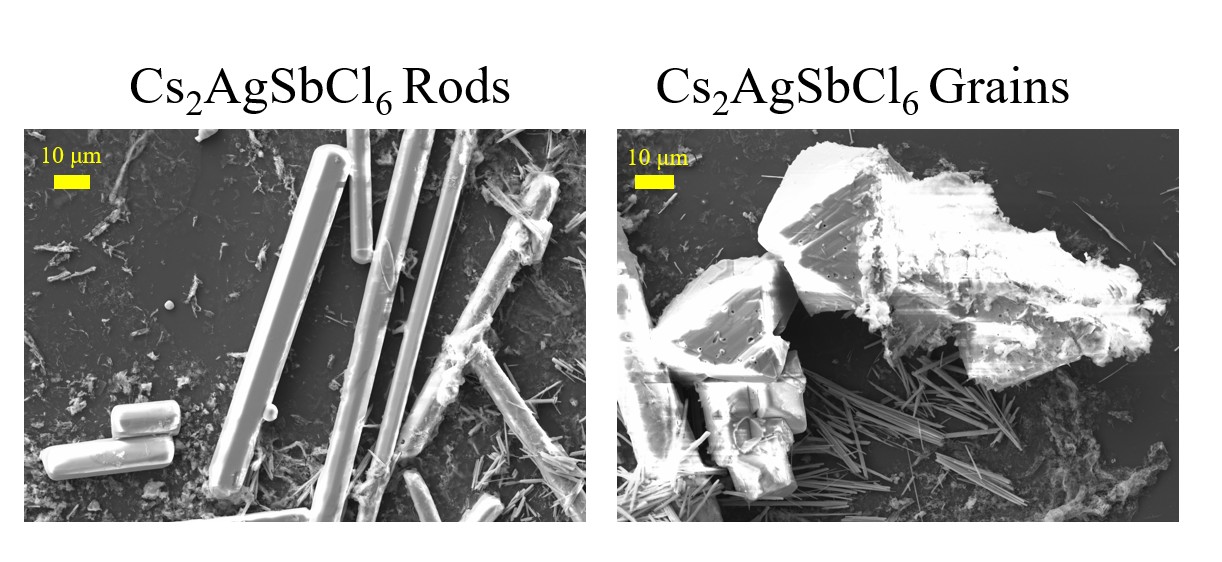}
    
    \caption{SEM images of the rods and grains of the post in-situ XRD annealed $Cs_2AgSbCl_6$. These images were taken before EDS spectra was taken.}
\end{figure}

\end{document}